# Trends in structural, electronic properties, Fermi surface topology, and inter-atomic bonding in the series of ternary layered dichalcogenides $KNi_2S_2$, $KNi_2Se_2$, and $KNi_2Te_2$ from first principles calculations

V. V. Bannikov and A. L. Ivanovskii *
*Institute of Solid State Chemistry, Ural Branch, Russian Academy of Sciences,
Pervomaiskaya St., 91, Ekaterinburg, 620990 Russia*
*E-mail address:* ivanovskii@ihim.uran.ru

A B S T R A C T

By means of the FLAPW-GGA approach, we have systematically studied the structural and electronic properties of tetragonal dichalcogenides $KNi_2Ch_2$ ($Ch$ = S, Se, and Te). Our results show that replacements of chalcogens (S → Se → Te) lead to anisotropic deformations of the crystals structure, which are related to the strong anisotropic character of the inter-atomic bonds, where inside the [$Ni_2Ch_2$] blocks, mixed covalent-ionic-metallic bonds occur, whereas between the adjacent [$Ni_2Ch_2$] blocks and K atomic sheets, ionic bonds emerge. We found that in the sequence $KNi_2S_2$ → $KNi_2Se_2$ → $KNi_2Te_2$ (i) the overall band structure (where the near-Fermi valence bands are due mainly to the Ni states) is preserved, but the width of the common valence band and the widths of the separate subbands and the gaps decrease; (ii) the total DOSs at the Fermi level also decrease; and (iii) for the Fermi surfaces, the most appreciable changes are demonstrated by the hole-like sheets, when a necklace-like topology is formed for the 2D-like sheets and the volume of the closed pockets decreases. Some trends in structural and electronic parameters for $ThCr_2Si_2$-type layered dichalcogenides, $KNi_2Ch_2$, $KFe_2Ch_2$, $KCo_2Se_2$, are discussed.

## 1. Introduction

The recent discovery [1] of a new family of $ThCr_2Si_2$-type layered Fe-$Ch$ superconductors (where $Ch$ are chalcogens), for which superconductivity emerges in [$Fe_2Ch_2$] blocks, has stimulated comprehensive investigations of their properties as well as the search for new related layered materials, see reviews [2-4]. Herein, extensive efforts, both experimental and theoretical, are directed at the development of new layered Fe-free ternary dichalcogenides. Very recently, the layered dichalcogenides $KNi_2Se_2$ and $KNi_2S_2$ (which exhibit a set of exciting physical properties such as unusual structural transitions, characteristics of the heavy-fermion behavior, low-temperature superconductivity, *etc*.) attracted much attention [5-9].

In this work, in order to get a systematic insight into the general trends in the basic structural, electronic properties, Fermi surface topology, and the peculiarities of inter-atomic bonding in Ni-based layered dichalcogenides depending on anion ($Ch^{2-}$) substitutions, we carried out a first-principles study of three related phases, namely, the synthesized $KNi_2S_2$ and $KNi_2Se_2$, as well as the hypothetical phase

KNi$_2$Te$_2$. Our results cover the optimized lattice parameters and atomic positions, electronic bands, Fermi surface topology, as well as total and partial densities of electronic states. Besides, Bader's analysis and the charge density maps are used in discussing the chemical bonding for the examined materials.

## 2. Models and computational aspects

We have examined the ternary dichalcogenides with the nominal composition KNi$_2$$Ch$$_2$ ($Ch$ = S, Se, and Te) and tetragonal ThCr$_2$Si$_2$-type structure (space group $I4/mmm$; #139, $Z$ = 2), Fig. 1. Here, the Ni atoms form a square lattice, whereas the $Ch$ atoms are located at the apical sites of the tetrahedrons {Fe$Ch$$_4$}. In turn, these tetrahedrons form quasi-two-dimensional (2D) blocks [Ni$_2$$Ch$$_2$]. In general, the structure of KNi$_2$$Ch$$_2$ can be schematically described as a stacking of potassium atomic sheets and [Ni$_2$$Ch$$_2$] blocks in the sequence: …[Ni$_2$$Ch$$_2$]/K/[Ni$_2$$Ch$$_2$]/K… The atomic positions are K: 2$a$(0,0,0), Ni: 4$d$(0,½,¼), and $Ch$: 4$e$(0,0,$z_{Ch}$), where $z_{Ch}$ are so-called internal coordinates governing the Ni - $Ch$ distances and the distortion of the Ni$Ch$$_4$ tetrahedra in the [Ni$_2$$Ch$$_2$] blocks.

All the calculations were carried out by means of the full-potential method within mixed basis APW+lo (LAPW) implemented in the WIEN2k suite of programs [10]. The generalized gradient correction (GGA) to exchange-correlation potential in the PBE form [11] was applied. The plane-wave expansion was taken to $R_{MT} \times K_{MAX}$ equal to 7, and the $k$ sampling with 10×10×10 $k$-points in the Brillouin zone was used. In all the calculations, we made full structural optimization of the ideal $A$Fe$_2$$Ch$$_2$ phases both over the lattice parameters and the atomic positions including the internal coordinates. The self-consistent calculations were considered to be converged when the difference in the total energy of the crystal did not exceed 0.1 mRy and the difference in the total electronic charge did not exceed 0.001 $e$ as calculated at consecutive steps.

Our results cover a representative set of physical parameters of KNi$_2$$Ch$$_2$ such as optimized atomic positions and lattice parameters, electronic bands, total and partial densities of states (DOSs), and Fermi surfaces. In addition, the bonding picture for KNi$_2$$Ch$$_2$ phases was discussed. For this purpose, the hybridization effects (covalent bonding) were analyzed using the densities of states (which were obtained by the modified tetrahedron method [12]) and were visualized by means of charge density maps. In turn, to estimate the amount of electrons redistributed between various atoms (ionic bonding), we carried out a Bader [13] analysis.

## 3. Results and discussion

### 3.1. Structural parameters

At the first step, the optimized structural parameters (including the internal coordinate $z_{Ch}$) are determined for the KNi$_2$$Ch$$_2$ phases; the calculated values (see the Table) are in reasonable agreement with the available data [5-9]. Some deviations of our results from the experimental data for KNi$_2$S$_2$ and KNi$_2$Se$_2$ may be related mainly to the deviation of the synthesized samples from the "ideal" stoichiometry 1:2:2 owing to cation deficiency [5-8], as well as to the well-known overestimation of the lattice parameters within LDA-GGA based calculation methods.

Our results show that in the sequence KNi$_2$S$_2$ → KNi$_2$Se$_2$ → KNi$_2$Te$_2$, both the *a* and *c* parameters increase, and this result can be easily explained by considering the atomic radii of *Ch* atoms: $R^{atom}$ (S) = 1.22 Å < $R^{atom}$ (Se) = 1.60 Å < $R^{atom}$ (Te) = 1.70 Å. At the same time, when going from KNi$_2$S$_2$ to KNi$_2$Te$_2$, the lattice expansion along *z* axis (growth of the parameter c) becomes much higher than the expansion in *xy* plane (growth of the parameter *a*): $\delta c = \{c^{KNi2Te2} - c^{KNi2S2}\}/c^{KNi2S2}$ = 14.3% *versus* $\delta a = \{a^{KNi2Te2} - a^{KNi2S2}\}/a^{KNi2S2}$ = 6.4%. Thus, the replacements of the chalcogens lead to *anisotropic deformations* of the crystal structure caused by strong anisotropy of inter-atomic bonds (see below). A similar effect was also found for a series of related 122 phases, see [14-16].

*3.2. Electronic structure and Fermi surfaces*

The energy bands, Fermi surfaces (FSs), as well as densities of states (DOS) for the examined KNi$_2$*Ch*$_2$ phases, which have been calculated for their equilibrium geometries, are shown in Figs. 2-5, respectively.

As can be seen, their electronic spectra are quite similar. Therefore, we will discuss their peculiarities in more detail using KNi$_2$Se$_2$ as an example. In Fig. 2, the near-Fermi band structure of KNi$_2$Se$_2$ is depicted along the selected high-symmetry lines within the first Brillouin zone of the body-centered tetragonal crystal. Here, the group of low-lying bands with the top at -3.0 eV below the Fermi level (E$_F$) is separated by a gap (~0.4 eV) from the partially occupied near-Fermi bands, which are located in the energy range from -2.6 eV to + 0.8 eV. There is no gap at E$_F$ providing metallic-like conductivity of KNi$_2$Se$_2$. In turn, these bands are separated from the next group of unoccupied bands by a gap at ~ 0.8 eV. From the atomic-resolved DOSs (Fig. 5) we conclude that the low-lying bands are composed mainly of hybridized valence states of Se and Ni, forming directional covalent bonds inside blocks [Ni$_2$Se$_2$] (see Fig. 1), whereas the bands in the window around the Fermi level are mainly of the Ni - character.

The near-Fermi bands demonstrate a complicated "mixed" character, combining the quasi-flat bands with a series of high-dispersive bands intersecting the Fermi level, Fig. 2. These features yield a multi-sheet FS, which consists (Fig. 3) of two disconnected electronic-like quasi-two-dimensional sheets (in the form of deformed cylinders extended along the $k_z$ direction), which are located around the corners of the Brillouin zone. Besides, two hole-like sheets are present. One of them is also 2D-like and is located on the lateral sides of the Brillouin zone, the other is closed, has a complicated form, and is centered at the *Γ* point, Fig. 3.

In the sequence $KNi_2S_2 \to KNi_2Se_2 \to KNi_2Te_2$, the hole-like sheets demonstrate the most appreciable changes, when a necklace-like topology is formed for the 2D-like sheets and the volume of the closed pockets decreases.

From Fig. 4 we see that in the sequence $KNi_2S_2 \to KNi_2Se_2 \to KNi_2Te_2$, the overall DOS structure is preserved, but the width of the common valence band and the widths of the subbands A and B and the gaps decrease. Besides, obvious changes occur in the total DOSs at the Fermi level, $N(E_F)$, which decrease as going from $KNi_2S_2$ to $KNi_2Te_2$, see the Table. This trend coincides with the experimental results [5,6], where the Sommerfield coefficient (γ, extracted from low-temperature specific heat data and related to the total density of states at the Fermi level) was found to decrease in the sequence $KNi_2S_2 \to KNi_2Se_2$.

Finally, let us summarize (Fig. 6) some trends in the structural and electronic parameters for $ThCr_2Si_2$-type layered dichalcogenides using the data for $KNi_2Ch_2$, as well as the results for $KFe_2Ch_2$ and $KCo_2Se_2$, which were obtained earlier within the same non-magnetic FLAPW-GGA calculations [16,17]. We see that for all dichalcogenides $KM_2Ch_2$ the main trends in changes of the lattice parameters depending on the anion type ($Ch$ = S, Se, and Te) coincide: (i). at the replacement of chalcogens (S → Se → Te) the both lattice parameters increase, and (ii). *anisotropic deformations* of the crystals occur. On the contrary, in this sequence, the changes in $N(E_F)$ depending on the cation type demonstrate the inverse trends: the values of $N(E_F)$ increase for $KFe_2Ch_2$ and decrease for $KNi_2Ch_2$, Fig. 6. Thus, we can speculate that further fine tuning of the electronic system of $KM_2Ch_2$ materials can be achieved using the co-doping strategy, *i.e.* by simultaneous substitutions in the cation and anion sublattices.

*3.3. Inter-atomic bonding*

To describe the inter-atomic bonding for the examined $KNi_2Ch_2$ phases, it is convenient to begin with a standard ionic picture, which considers the usual oxidation numbers of atoms: $K^{1+}$, $Ni^{2+}$, and $Ch^{2-}$. Thus, the formal charge state (in ideal stoichiometry 1:2:2) for nickel should be 1.5+. In this approximation, the "universal" ionic formula for the examined phases should be $K^{1+}[Ni^{1.5+}_2 Ch^{2-}_2]^{1-}$. This means identical atomic charge states and identical inter-blocks charge transfer ($1e$) irrespective of the chemical composition of $KNi_2Ch_2$ phases.

Unlike this over-simplified ionic picture, our calculations reveal that the inter-atomic interactions for these phases can be described as a high-anisotropic mixture of metallic, covalent, and ionic contributions.

Herein, the *metallic-like* bonding is due to the interaction of the near-Fermi Ni states inside blocks [$Ni_2Ch_2$]. From the Table we see that the values of $N(E_F)$ decrease in the sequence $KNi_2S_2 \to KNi_2Se_2 \to KNi_2Te_2$, it means the weakening of these *metallic-like* interactions.

Next, *covalent* bonds also arise inside blocks [$Ni_2Ch_2$] and may be well understood from the charge density maps, Fig. 1. It is seen that together with the strongest covalent bonds Ni-*Ch*, some covalent *Ch-Ch* bonding occurs. Certainly, these covalent bonds decrease as going from KNi$_2$S$_2$ to KNi$_2$Te$_2$ as a result of the bond length growth.

As to *ionic* bonding, the Bader's model [13] was used for estimating the actual atomic charges. The calculated atomic charges (K$^{0.801+}$Ni$^{0.466+}_2$S$^{0.866-}_2$, K$^{0.787+}$Ni$^{0.284+}_2$S$^{0.677-}_2$, and K$^{0.772+}$Ni$^{\sim 0.010+}_2$S$^{0.396-}_2$) differ from the formal ionic charges owing to covalence in blocks [Ni$_2$*Ch*$_2$], and in the sequence KNi$_2$S$_2$ → KNi$_2$Se$_2$ → KNi$_2$Te$_2$, the charge transfer between the adjacent atomic K sheets and the blocks [Ni$_2$*Ch*$_2$] decreases. Thus, in this sequence, the chemical bonding weakens owing to reduction of the covalent and ionic contributions.

## 4. Conclusions

In summary, by means of the FLAPW-GGA approach, we have systematically studied the structural and electronic properties of a group of tetragonal ternary dichalcogenides KNi$_2$*Ch*$_2$ (*Ch* = S, Se, and Te).

Our results show that replacements of chalcogens (S → Se → Te) lead to *anisotropic deformations* of the crystals structure caused by strong anisotropy of inter-atomic bonds.

We found that in the sequence KNi$_2$S$_2$ → KNi$_2$Se$_2$ → KNi$_2$Te$_2$, (i) the basic band structure picture is preserved, but the width of the common valence band and the widths of the separate subbands and the gaps decrease; (ii) the total DOSs at the Fermi level also decrease; and (iii) for the FSs, the most appreciable changes are demonstrated by the hole-like sheets, when a necklace-like topology is formed for the 2D-like sheets and the volume of the closed pockets decreases.

Finally, our analysis reveals that the inter-atomic bonding for KNi$_2$*Ch*$_2$ is of a high-anisotropic character: inside the [Ni$_2$*Ch*$_2$] blocks, mixed covalent-ionic-metallic bonds occur, whereas between the adjacent [Ni$_2$*Ch*$_2$] blocks and K atomic sheets, ionic bonds emerge. Thus, the examined phases may be classified as *ionic metals*. Finally, we found that in the sequence KNi$_2$S$_2$ → KNi$_2$Se$_2$ → KNi$_2$Te$_2$ the chemical bonding weakens.

**Table**

The optimized lattice parameters ($a$, $c$, in Å), $c/a$ ratio, internal coordinates $z_{Ch}$, and total density of states at the Fermi level ($N(E_F)$, in states/(eV·cell)) for $KNi_2S_2$, $KNi_2Se_2$, and $KNi_2Te_2$ phases as obtained within FLAPW-GGA.

| phase | $a$ | $c$ | $c/a$ | $z_{Ch}$ | $N(E_F)$ |
|---|---|---|---|---|---|
| $KNi_2S_2$ | 3.81 (3.779 [1]; 3.7792 *) | 12.36 (12.714 [1]; 12.7139 *) | 3.244 (3.364 [1]) | 0.350 (0.350 [1]; 0.35*) | 7.322 (11.5 LDA+G*) |
| $KNi_2Se_2$ | 3.94 (3.893 [2]; 3.910 [3]; 3.899 [4]; 3.968 [5]) | 12.95 (13.316 [2]; 13.414 [3]; 13.473 [4]; 13.048 [5]) | 3.287 (3.421 [2]; 3.431 [3]; 3.456 [4]; 3.288 [5]) | 0.354 (0.355 [2]; 0.354 [3]; 0.351 [5]) | 6.303 |
| $KNi_2Te_2$ | 4.13 | 14.13 | 3.421 | 0.355 | 5.687 |

The available data are given in parentheses.
[2] Ref. [5], experimental
[2] Ref. [6], experimental
[3] Ref. [7], experimental
[4] Ref. [8], experimental
[5] Ref. [9], calculated; DFT-GGA
* Ref. [18], calculated; LDA/LDA+G

**FIGURES**

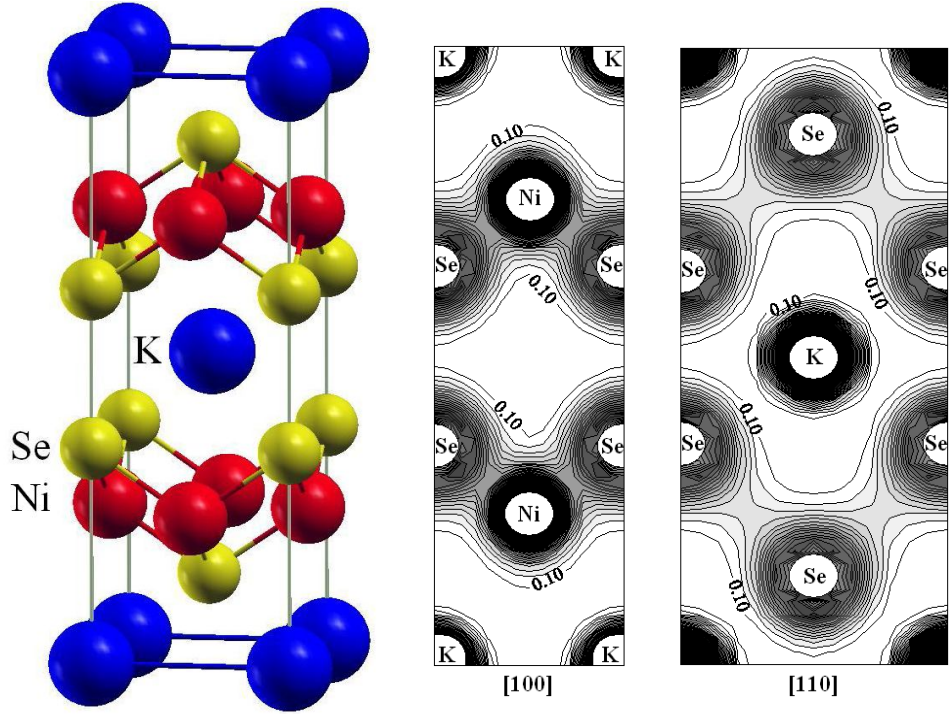

**Fig. 1.** (*Color online*). Crystal structure of the tetragonal ThCr$_2$Si$_2$-type KNi$_2$Se$_2$ phase and the valence charge density maps in [100] and [110] planes. The distance between the ρ contours is 0.05 $e$/Å$^3$.

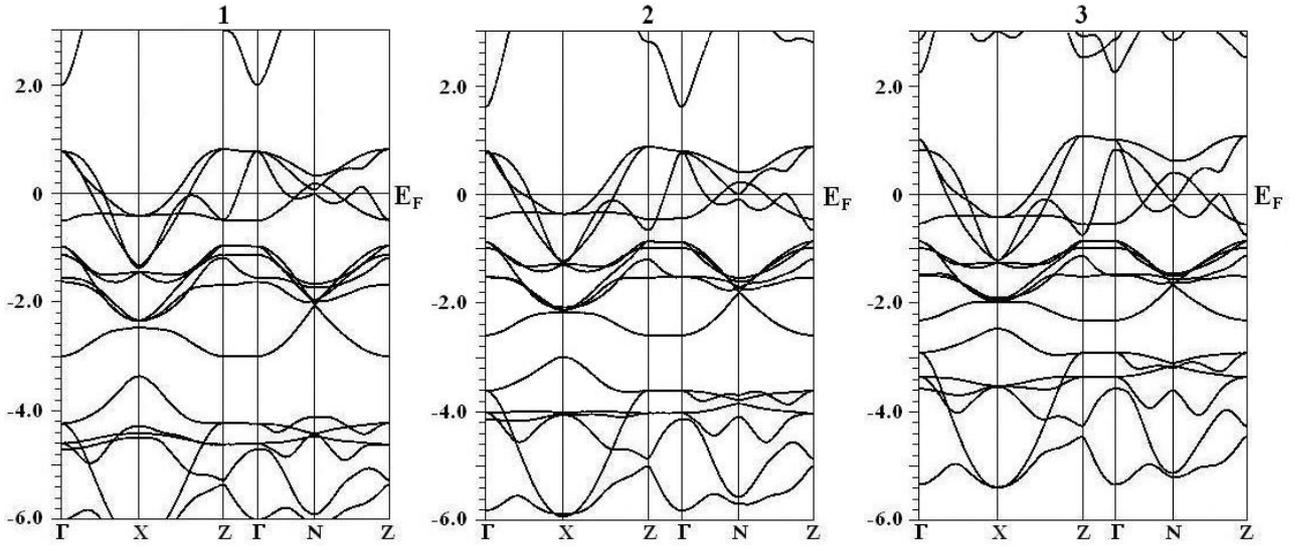

**Fig. 2.** Electronic band structures of KNi$_2$S$_2$ (1), KNi$_2$Se$_2$ (2), and KNi$_2$Te$_2$ (3).

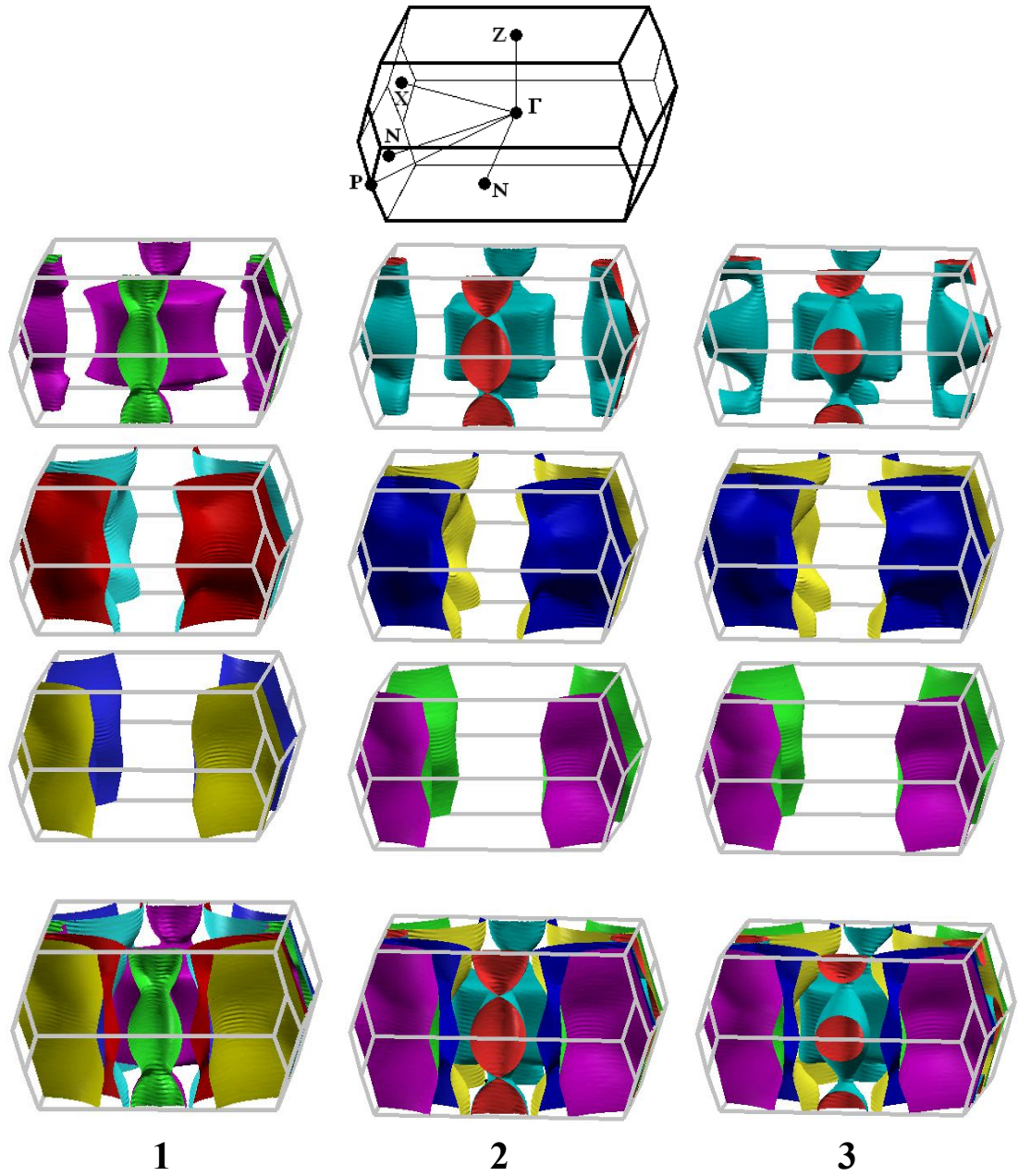

**Fig. 3.** (*Color online*) The Fermi surfaces of KNi$_2$S$_2$ (1), KNi$_2$Se$_2$ (2), and KNi$_2$Te$_2$ (3).

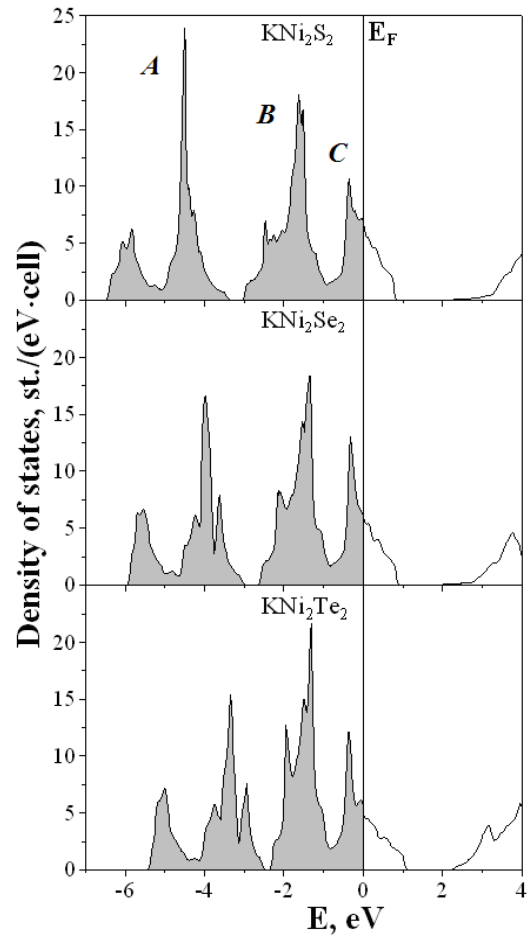

**Fig. 4.** Total densities of states of KNi$_2$S$_2$, KNi$_2$Se$_2$, and KNi$_2$Te$_2$.

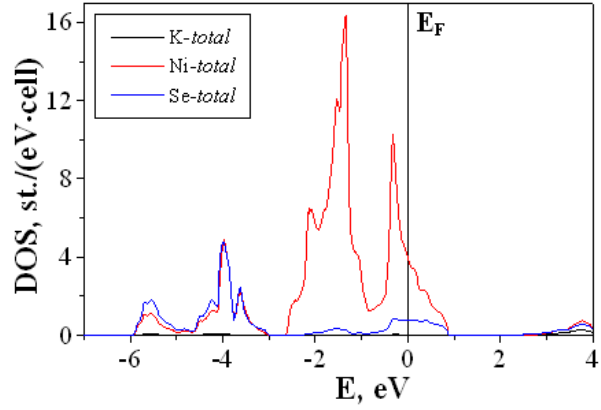

**Fig. 5.** (*Color online*) Atomic-resolved DOSs for KNi$_2$Se$_2$.

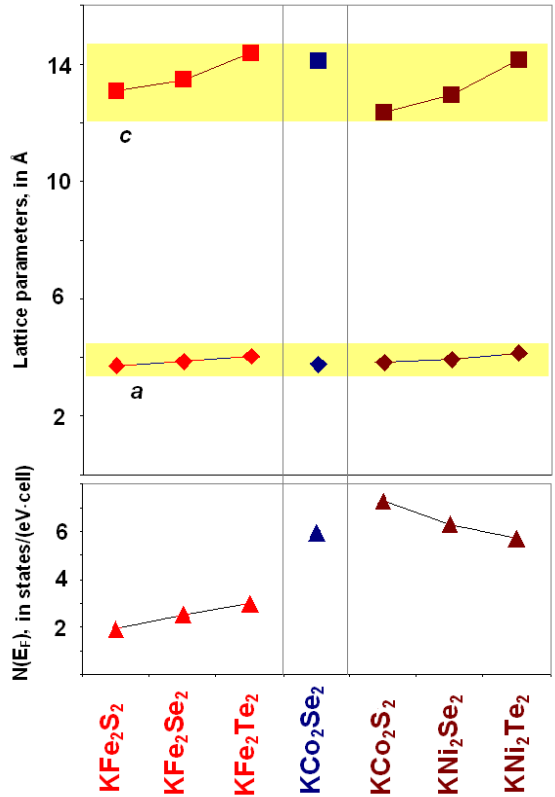

**Fig. 6.** Changes in the lattice parameters (*a* and *c*) and the total density of states at the Fermi level ($N(E_F)$) for the series of ThCr$_2$Si$_2$-type layered dichalcogenides K$M_2Ch_2$ depending on the cation (*M*= Fe [16], Co [17], and Ni) and anion types (*Ch* = S, Se, and Te) as calculated within FLAPW-GGA.